\documentclass[prd,aps,twocolumn,showpacs,preprintnumbers,amsmath,amssymb]{revtex4}
\usepackage[dvips]{graphicx} 
\usepackage{amsmath} 
\usepackage{amssymb} 
\usepackage{graphicx}
\usepackage{dcolumn}
\usepackage{bm}
\def\be{\begin{equation}} 
\def\ee{\end{equation}} 
\def\bea{\begin{eqnarray}} 
\def\eea{\end{eqnarray}} 
 
\begin{document} 
 
 
\date{\today} 
 
\title{Detecting Cosmic Strings in the CMB with the Canny Algorithm} 
 
\author{Stephen Amsel$^{1)}$ \email[email: ]{samsel@po.box.mcgill.ca}}

\author{Joshua Berger$^{2)}$ \email[email: ]{jb454@cornell.edu}} 

\author{Robert H. Brandenberger$^{1)}$ 
\email[email: ]{rhb@hep.physics.mcgill.ca}}
 
\affiliation{1) Department of Physics, McGill University, 
Montr\'eal, QC, H3A 2T8, Canada} 
 
\affiliation{2) Department of Physics, Cornell University,
Ithaca, NY, , USA}

\pacs{98.80.Cq} 
 
\begin{abstract} 

Line discontinuities in cosmic microwave background anisotropy maps
are a distinctive prediction of models with cosmic strings. These
signatures are visible in anisotropy maps with good angular resolution
and should be identifiable using edge detection algorithms. One such
algorithm is the Canny algorithm. We study the potential of this
algorithm to pick out the line discontinuities generated by cosmic strings. 
By applying the algorithm to small-scale microwave anisotropy maps
generated from theoretical models with and without cosmic strings, we
find that, given an angular resolution of several minutes of arc, 
cosmic strings can be detected down to a limit of the mass per unit length
of the string which is one order of magnitude lower than the current
upper bounds.
 
\end{abstract} 
 
\maketitle

\newcommand{\eq}[2]{\begin{equation}\label{#1}{#2}\end{equation}} 
 
\section{Introduction} 

Cosmic strings \cite{original} are one-dimensional topological defects 
which arise during phase transitions in the very early universe. Since
they carry energy, they will lead to density fluctuations and
cosmic microwave background (CMB) anisotropies (see e.g.
\cite{CSreviews} for reviews on cosmic strings and structure formation).
Causality implies that the network of strings which forms during the
phase transition contains infinite strings.
Once formed in the early universe, the network of strings will approach a
``scaling solution'' which is characterized by of the order one infinite
string segment in each Hubble volume, and a distribution of cosmic string
loops which are the remnants of the previous evolution. In particular,
this implies that in any theory which admits cosmic strings, a network of
strings will be present during the time period relevant to the CMB,
namely between the time of recombination $t_{rec}$ and the present time
$t_0$. 

A network of cosmic strings will generate a scale-invariant spectrum of
cosmological perturbations \cite{CSstructure}. As first discussed
by Kaiser and Stebbins (KS) \cite{KS}, 
the non-Gaussian nature of the density field
produced by strings will lead to a distinctive signature 
(which we will call KS signature) in CMB anisotropy
maps, namely line discontinuities. These line discontinuities are a
consequence of the non-trivial nature of the metric produced by a
cosmic string: space perpendicular to a cosmic string is a cone with
deficit angle given by
\be 
\alpha \, = \, 8 \pi G \mu \, ,
\ee
where $\mu$ is the mass per unit length of the string and $G$ is Newton's
gravitational constant \cite{deficit}. Since cosmic strings have a
tension comparable to their mass per unit length, they will typically
be moving with a relativistic transverse velocity $v$. As illustrated in
Figure 1, if we are looking at the CMB in direction of the string, we
will see the photons passing on different sides of the string with a
Doppler shift
\be \label{KSsig}
{{\delta T} \over T} \, = \, 8 \pi \gamma(v) v G \mu \, ,
\ee
where $\gamma(v)$ is the relativistic gamma factor. Looking in direction
of the string, we will see a line in the sky across which the CMB
temperature jumps by the above amount. 

\begin{figure}
\includegraphics[height=6cm]{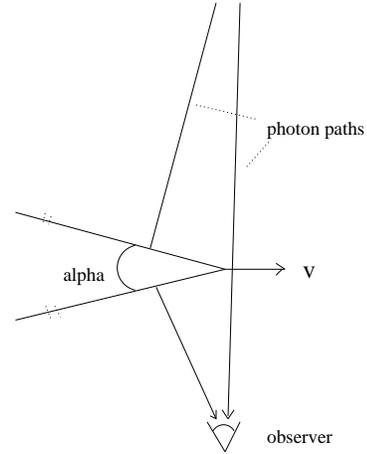}
\caption{Geometry of the Kaiser-Stebbins effect: Photons passing
on the two sides of the moving cosmic string obtain a relative Doppler
shift for the observer who is at rest.} \label{fig:1}
\end{figure}

Any cosmic string which photons
reaching us today pass on their way from the last scattering surface will
produce an effect. The most numerous strings, however, are those present
close to the time of last scattering $t_{rec}$. At that time, the
typical curvature radius of a long string segment is of the order of
the Hubble radius at $t_{rec}$. The corresponding angular scale today
is about $1^{0}$. In order to be able to identify the cosmic signature
as a line discontinuity, as good an angular resolution as possible is 
required. If probed with an angular resolution not significantly
smaller than $1^{0}$, the KS signature will be washed out.
In this work we will study the potential of surveys with
an angular of several minutes of arc to detect the KS signature for
strings. Less important than the angular resolution of the survey is
the total sky coverage, as long as it includes a sufficiently
large number of Hubble patches at $t_{rec}$. Obviously, increasing the
sky coverage will reduce the standard deviation of the results and
thus lead to some better discriminating power. 
  
There has been surprisingly little work devoted to detecting the KS
signature. An early study \cite{Moessner} showed that the angular
resolution of the WMAP satellite would not be adequate to pick out
the KS signature, even for values of $G \mu$ for which the cosmic
strings would dominate the power spectrum of density fluctuations.
In order to detect the KS signature, new statistical methods must
be employed. In \cite{Lo}, a matched filtering method was applied
to the WMAP data to search for cosmic string signatures. In
\cite{Smoot}, new statistics (such as a statistic measuring
the connectedness of neighboring temperature steps or
a decomposition of the temperature map into constant background
temperature, Gaussian noise plus straight string step
discontinuities) were introduced 
and applied to the WMAP
data to search for strings. Based on the null results of both studies,
an upper bound on $G \mu$ of $G \mu < 10^{-6}$ could be set.
However, this bound is not competitive with other existing bounds on
the string tension coming from the observed acoustic peak structure
of the CMB angular power spectrum \cite{CMBlimit}
(which yields $G \mu < 10^{-7}$) and from pulsar timing measurements
\cite{Pulsarlimit} (which yield limits of between 
$G \mu < 2 \times 10^{-8}$ and $G \mu < 10^{-5.5}$, the difference being 
due to conflicting ways \cite{timinglimits}
in which the pulsar timing data is analyzed to obtain limits on
the amplitude of the stochastic background of gravitational waves, and due
to different assumptions about the distribution of cosmic string loops,
the main source of gravitational radiation from a string network -
see \cite{Pol} for a recent discussion). Note that an advantage
of using the KS signature over pulsar bounds is that the KS signature
is more robust (less dependent on the unknown distribution of cosmic
string loops) than the pulsar bounds (which depend sensitively on
the details of the cosmic string loop distribution)
\footnote{The statistical
properties of the long strings are much better established from
numerical string evolution simulations than is the distribution of 
string loops. Different numerical codes all agree that the 
distribution of long strings scales. Even the number of long
strings crossing each Hubble volume is known to 
within one order of magnitude. The distribution of cosmic string loops,
however, much less well known. Even the form of the scaling distribution,
not to mention the parameters of such a distribution, are unknown.}

As discussed above, we expect to be able to set much stronger limits
on $G \mu$ from searches for the KS signature by looking at small-scale
anisotropy maps, provided we find a good statistic to identify line
discontinuities. The goal of this paper is to explore the potential
of one specific edge detection technique, the Canny algorithm \cite{Canny}
to pick out these line discontinuities in CMB anisotropy maps.

One reason for the relative lack of previous work on detecting the
KS signature of cosmic strings is the fact that interest in cosmic
strings as a source for structure in the universe \footnote{Initially
\cite{CSstructure}, cosmic strings were studied as an alternative to
inflation as a mechanism for structure formation.} decreased
dramatically after the discovery of the sharpness of the acoustic peaks
in the CMB angular power spectrum \cite{Boomerang,WMAP}. Since
perturbations seeded by strings are not coherent \cite{incoherent},
a scenario with only cosmic strings as the source of density perturbations
would only produce one broad Doppler peak rather than the observed
narrow acoustic oscillations \cite{Periv,Albrecht,Turok}. 

However, recent years have seen a resurgence of interest in cosmic strings.
This was fueled by two developments. Firstly, it was realized \cite{Rachel}
that in many supersymmetric particle physics models, cosmic strings are
formed after inflation, and thus contribute to but do not completely
replace inflationary perturbations as the seeds for structure formation.
Secondly, it has recently been realized that models with cosmic superstrings 
\cite{Witten} may well be viable \cite{CMP}. They could, for example,
be generated as the remnant of brane annihilation processes in
brane inflation models \cite{Tye}, or they may play an important roles
in inflationary models in warped backgrounds \cite{stringinflation}.
Cosmic superstrings may also be left behind after the initial Hagedorn
phase in string gas cosmology \cite{BV}, where they would add an additional
component to the spectrum of fluctuations produced by thermal string
gas fluctuations \cite{NBV}. Thus, it is of great interest to
find new ways to search for signatures of strings in cosmological data.

The outline of this paper is as follows: In the following
section we review the Canny algorithm and describe its application
to our problem. Section 3 discusses how the temperature maps 
with and without strings are produced, introduces the parameters
chosen in the specific simulations, and presents our results.
We conclude with a summary and a discussion of the caveats
of the current analysis and prospects for future work.

\section{The Canny Algorithm}

The Canny algorithm was developed in 1986 as a technique to detect
edges in images \cite{Canny} such as the
two-dimensional images considered here. When applied to a map of raw data,
the algorithm is intended to produce a map tracing the edges in the
map, the lines across which the intensity contrast is largest.

The first step in the algorithm is to filter the data to eliminate
point source noise. The filtering is achieved using a convolution of
the map with a Gaussian filter. The filtering length is a free parameter
in the algorithm. It must be chosen sufficiently large to eliminate
unwanted point source noise, but must be smaller than the characteristic
size of the structure in the maps which one is trying to identify. Let us
denote the original map by $M(i, j)$, and the filter by $F(i, j)$. The
filtered map $FM$ is then given by
\be
FM(i, j) \, = \, \sum_{k,l} M(i - k, j - l) F(k, l) \, ,
\ee
where the pixel points of the map are denoted by the labels $(i, j)$. 
In our simulations, the filtering length is taken to be $1.5$ grid
units.

The second step of the algorithm is to find the gradient of the
filtered image, a vector obtained by taking the discrete derivative
of the map in each of the two directions. The gradient vector at
a grid point $(i, j)$ is denoted by $(G_x, G_y)$, where $G_x$ and $G_y$
are the derivatives in the respective directions. The {\it edge
strength} $|G|$ at any given grid point is defined by
\be
|G| \, = \,\sqrt{|G_x|^2 + |G_y|^2} \, ,
\ee
and the {\it edge direction} is given by the angle
\be
\theta \, = \, arctan(G_y / G_x) \, .
\ee
Since we are working on a grid, it makes sense to replace the edge
direction by one of the eight distinguished directions on a quadratic
grid, namely the four directions along the coordinate axes and the
four diagonal directions.

The average maximal gradient in maps without cosmic strings (the
average taken over all of our runs) is denoted by $G_{m}$. 
For a specific run, the maximal gradient is obtained by scanning
the entire map and searching for the maximal value of the grid
strength. The value of $G_{m}$ is
used to set the thresholds discussed below.

The next step of the Canny algorithm is to find grid points with 
gradients which are maximal when we vary the point
in direction of the gradient. Grid
points which are not local maxima of the gradient are assigned
a number $0$. 

At this stage, two thresholds must be set, an upper and a lower 
threshold $t_u$ and $t_l$. The algorithm first finds local maxima in 
the gradient along one of the four directional axes on the grid.  
Next, it goes through each of the list of local maxima and
determines whether the gradient is greater than the upper threshold fraction of $G_m$, i.e.
\be
|G| \, > \, t_u G_m \, .
\ee
Grid point thus selected are marked with a number $1$. If the
gradient is larger than the lower threshold fraction of $G_m$,
\be
|G| \, > \, t_l G_m \, ,
\ee
the grid point is assigned a value of $1/2$. For
points marked with $1$ or $1/2$, the direction of the gradient
is also stored. For each grid point with a value of $1/2$, the 
algorithm next checks whether there is a grid point with
value of $1$ or $1/2$ which is in a direction perpendicular
to the gradient and
whose gradient is parallel or next to parallel (thus three
possible directions in total) to the initial gradient. If such a
grid point is found, the algorithm continues until it either does
not find another point satisfying the criteria or else finds a
point satisfying the criteria marked with a $1$. If it finds such
a point, then all of the points (initially marked $1/2$) found along
the way are marked as $1$. Neighboring (in all eight directions)
grid points marked by $1$ are then said to belong to the same edge.
These are edges across which the gradient of the map is large.

The reason for using two thresholds is as follows: we want the algorithm
to find lines with a large gradient. At the same time, we do not want
noise to lead to points along a line of large gradients to be missed
(and thus the lines cut) just because noise has reduced the amplitude
of the edge strength at one point along the line.

In this way, the algorithm produces a list of grid points belonging to
edges, and this list can be read out as a map of edges.
In our simulations, the upper and lower thresholds are $t_u = 0.5$
and $t_l = 0.4$.

It is important to quantify the edge maps. A simple way to do this is
to produce a histogram of edge lengths. Thus, the implementation of
the Canny algorithm is arranged such as to output a list of edge
lengths which can then be used to produce a histogram of edge
lengths. The histogram contains useful information about the
presence of edges in the map.

\section{Applying the Canny Algorithm to CMB Anisotropy Maps}

Eventually, we would like to apply the Canny algorithm to actual
CMB maps. For this initial feasibility study, however, we will
apply the algorithm to theoretical maps produced in numerical
simulations. We wish to compare temperature maps for the
``standard'' Gaussian $\Lambda$CDM model with maps in which a network
of strings contributes a fraction $f$ of the total power.
The maps are characterized by the angular scale of the survey 
(we take the survey area to be square) and by the angular scale
of the grid, i.e. the angular resolution of the survey.

The Gaussian maps are produced in the following way: We start with
the angular power spectrum $C_l$ of cosmic microwave anisotropies
taken from the CMBFAST simulation \cite{CMBFAST}
with the appropriate cosmological parameters
\footnote{Since the range of $l$ values for which the CMBFAST
program generates $C_l$ values is limited to $3000$, we
are currently unable to explore angular scales relevant to
upcoming experiments. We are at the moment completing an improved
code which will allow us to obtain an improved angular resolution.}. 
Since we have
in mind applications of our algorithm to small angular scale surveys,
we perform a ``flat sky'' approximation \cite{White}. 
We introduce a two-dimensional
Cartesian coordinate system covering the survey area which we take
to be rectangular, choosing the upper right corner of the survey area
to be the origin of the coordinates. We need to compute the temperature
field $T_G(x, y)$ 
of the map at the coordinate values $(x, y)$ corresponding
to the grid points. This map is determined by an inverse fast Fourier
transform from the temperature values $\tilde{T}({\vec{k}})$ in
Fourier space.

For each vector ${\vec{k}}$ in Fourier space, we find the integers
$l_1(k)$ and $l_2(k)$ which bracket the $l$ value $l(k)$ corresponding
to ${\vec{k}}$ and take a linear interpolation of the values of $C_l$.
Let us denote the result of this linear interpolation by $C_{l(k)}$,
where $k \equiv |{\vec{k}}|$. The
value of ${\tilde{T}}({\vec{k}})$ is then given as follows:
\be
{\tilde{T}}({\vec{k}}) \, = \, g({\vec{k}}) {\sqrt{C_{l(k)}}} \, ,
\ee
where $g$ is a random variable drawn from a probability distribution
with variance $1$.

In the above analysis, we are singling out one grid point to be
special, namely the origin. This introduces unwanted phase
correlations which we overcome by superimposing the results from
four separate simulations, indicated by $T_i, i = 1 , .., 4$:
\bea
T_G(x, y) \, &=& \, {1 \over 2} \bigl( T_1(x, y) + T_2(x_m - x, y_m - y) \\
&& + T_3(x_m - x, y) + T_4(x, y_m - y) \bigr) \, , \nonumber
\eea
where $x_m$ and $y_m$ are the maximal $x$ and $y$ values of the survey
volume (the pre-factor of $1/2$ is required in order to maintain the
original standard deviation).
 
A network of cosmic strings will produce a temperature map
$T_{CS}(x, y)$. In the following, we take a toy model for a
temperature map produced by strings introduced by Perivolaropoulos
\cite{Leandros,Moessner}. The toy model takes into
account that at all times between $t_{rec}$ (the time of
recombination) and the present time $t_0$, the network of
strings is described by a scaling solution in which there are
a fixed number of long string segments (strings which are
not loops with radius smaller than the Hubble radius) crossing
each Hubble volume. Each such string gives rise to a 
Kaiser-Stebbins line discontinuity (\ref{KSsig})
in the temperature map. The cosmic string network is
continuously evolving via motion of the strings and string
interactions which lead to the production of string loops.
Thus, in each Hubble time step the string network can be
taken to be uncorrelated. 

In the algorithm, we divide the time interval $t_{rec} < t < t_0$
into 15 Hubble time steps. In each time step $t$ we lay down a network
of strings at random, uncorrelated with the network at the previous
time step. We take the network to consist of straight string
segments of length $\alpha_1 t$.  In order to avoid missing strings
at the edges of the survey area, the survey area is extended
in each direction by a Hubble distance. The program runs through
all points of the simulation volume and picks points to be
midpoints of a string segment with
a probability chosen such that the average number of strings in the
Hubble volume equals the number $N$ of the scaling solution. The
directions of the strings are chosen at random, as are the string
transverse velocities. To take into account also the projection onto 
the last scattering surface, we add a temperature
\be
{{\delta T} \over T} \, = \, {1 \over 2} {\tilde v} r 8 \pi G \mu \, , 
\ee
to one side of the string projection onto the last scattering surface, 
and subtract it from the other (so as to maintain the average temperature 
of the CMB). Here, ${\tilde v}$ stands for the maximal value of 
$v \gamma(v)$, where $v$ is the transverse velocity of the 
string, and $\gamma(v)$ is the relativistic $\gamma$ factor associated
with $v$. Also, $r$ is a random number between $0$ and $1$, to take
into account both the distribution of velocities, and also projection
effects (the formula (\ref{KSsig}) for the line
discontinuity in temperature is modulated by an angular factor if
the string velocity is not perpendicular to our line of sight
to the string).  The regions affected by the temperature fluctuation are
rectangles on either side of the string. The depth of these
rectangles in direction transverse to the direction of the string
is taken to be a fraction $f$ of the Hubble radius. The length
of the string segments is taken to be $\alpha_1$ times the
Hubble radius. The survey area
is always taken to correspond to the central region of the
simulation area.

The free parameters of the string simulation are $G \mu$, the
number of strings $N$ per Hubble volume, the length coefficients
$\alpha_1$ and $f$, as well as ${\tilde v}$. In our work, we
will fix $N = 10$, $\alpha_1 = 2$, $f = 1$ and ${\tilde v} = 0.15$.
These values are representative of the results of numerical
simulations of cosmic string evolution (see \cite{CSreviews} for reviews).
However, it should be kept in mind that the results for 
these parameters obtained using different numerical codes differ significantly. 

Let us make a couple of comments on the expected 
dependence of the results on the value of $N$. As $N$ initially
increases from zero, the signature of the strings should increase since
there are more strings. However, once $N$ rises above a certain
critical value, then further increasing the number $N$ will render the
string distribution more Gaussian by the Central Limit Theorem and
will hence decrease the sensitivity of the Canny algorithm. We
expect that the critical value of $N$ will be the one for which
the regions of the sky effected by individual string segments begin
to overlap. If the strings have length comparable to the Hubble radius,
the the critical value of $N$ is expected to be less than $N = 10$.
 
Increasing the string velocity ${\tilde v}$ will boost the line
discontinuites in the temerature maps and will make our algorithm
more effective. Since the weight in our statistical analysis of
the edge maps comes from fairly short segments, our results are
not very sensitive to the specific values of $f$ and $\alpha_1$.

We are interested in using the Canny algorithm to find signatures
for cosmic strings in maps which contain both Gaussian fluctuations
from inflation and a cosmic string component (which is sub-dominant
in terms of its contribution to the power spectrum). The total
map $T_T(x, y)$ is given by the superposition of a pure Gaussian
noise map and the map produced by a distribution of strings:
\be
T_T(x, y) \, = \, T_G(x, y) + T_{CS}(x, y) \, ,
\ee
and where the amplitude of the Gaussian term is adjusted
such that the total angular power spectrum agrees with the
COBE results \cite{COBE}.

The output of the Canny algorithm is a map of edges which have
been picked out. In the presence of cosmic strings, we would
expect more longer edges than in the absence of strings. In
order to test for this effect, we produced histograms of the
distribution of edge lengths for each simulation. Since both
the Gaussian maps and the maps with cosmic strings are produced
by a Gaussian random process (the phases of the Fourier modes for
the Gaussian maps are picked at random, and for the string maps
the locations of the string centers, their directions and their
transverse velocity vectors are all random), we ran many 
(${\tilde N} = 50$)
simulations for both the Gaussian map and the maps with strings.
This gave us statistical error bars on the histograms of edge
lengths. In turn, this allows us to assign statistical weight to
the difference in the histograms. Our results are based on this
analysis.

In our simulations, we consider the ``Gaussian noise'' maps to be
given by a standard $\Lambda$CDM model with adiabatic
fluctuations with a spectral index of $n_s = 0.99$, and
with parameters
$\Omega_B = 0.046$ (baryon fraction), $\Omega_{CDM} = 0.224$ (cold
dark matter fraction), $\Omega_{\Lambda} = 0.730$ (cosmological
constant contribution), and $H_0 = 72$ (Hubble expansion rate). No
massless neutrinos, standard recombination history and a Helium
fraction of $Y = 0.24$ were assumed (in the $C_l$ spectra which
were used to construct the maps). 

The parameters of the string simulation are indicated
above. Our maps have angular extent $15$ degrees by $15$ degrees.

We fix the angular resolution and search for the minimal value
of the cosmic string mass parameter $G \mu$ for which the maps
with the strings can be distinguished from the pure Gaussian
maps at a statistically significant level. We demand 
significance at the $3 \sigma$ level.

Figure 2 shows a CMB temperature fluctuation map of a simulation
without strings, and Figure 3 is the corresponding map of a simulation
which includes strings according to the prescription described above.

\begin{figure}
\includegraphics[height=9cm]{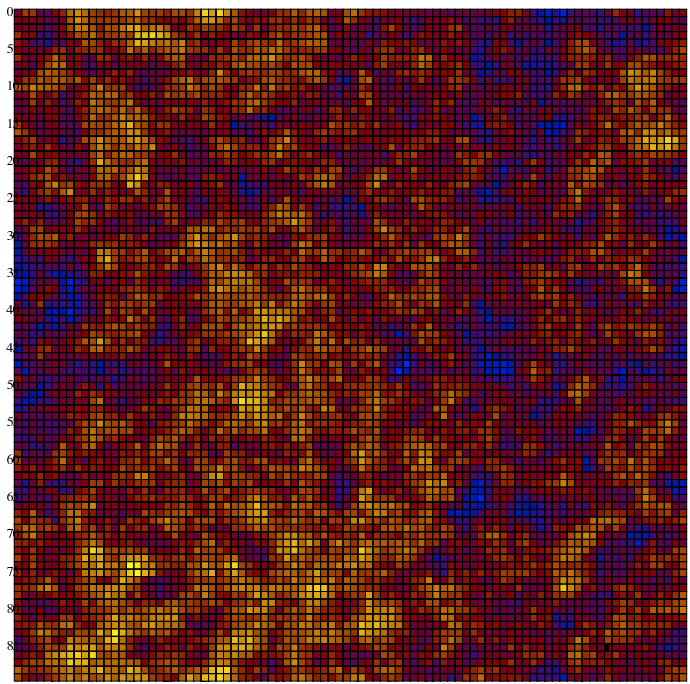}
\caption{Map of the CMB temperature in a $15^2$ square of the sky
for a $\Lambda$CDM simulation with the parameters described in the text.} 
\label{fig:2}
\end{figure}

\begin{figure}
\includegraphics[height=9cm]{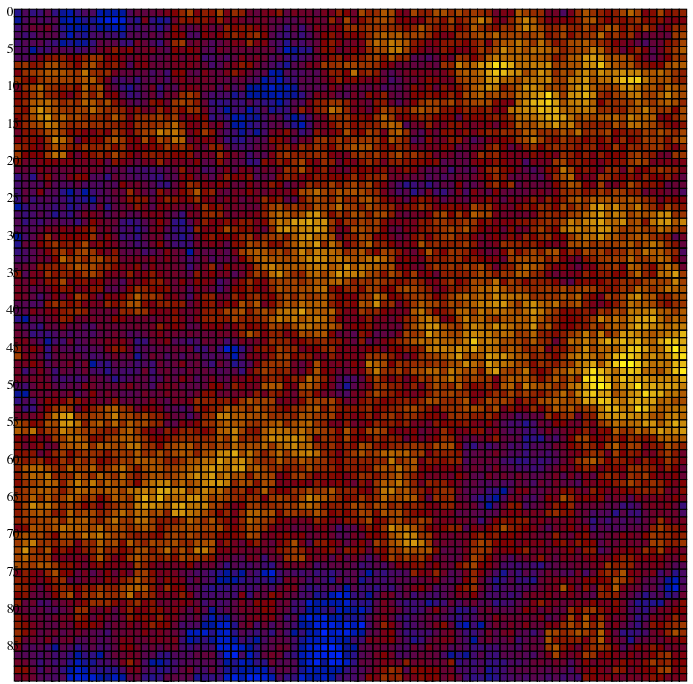}
\caption{Corresponding map of the CMB temperature in a simulation which
includes cosmic strings with a mass per unit length parameter given
by $G \mu = 10^{-7}$ (and other parameters as described in the text.} 
\label{fig:3}
\end{figure}

The output of the Canny algorithm applied to the above maps is
shown in Figures 4 and 5, respectively.

\begin{figure}
\includegraphics[height=9cm]{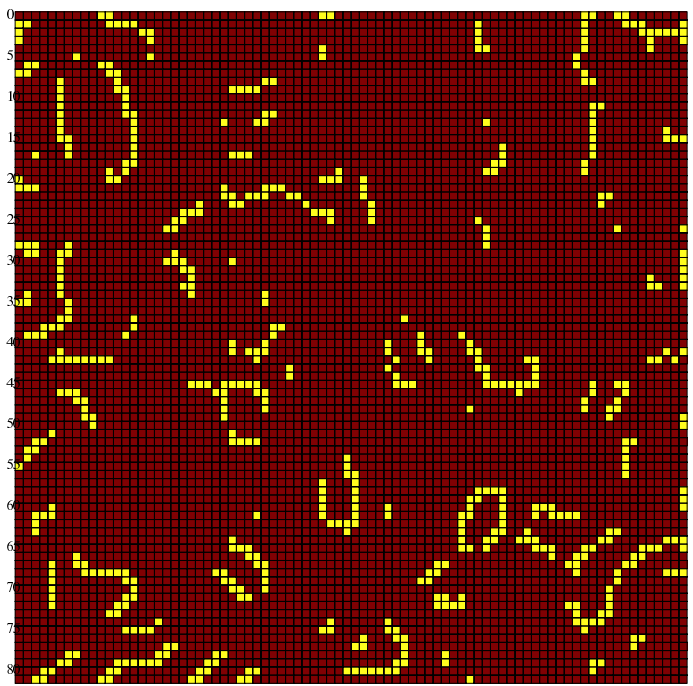}
\caption{Output map of the Canny algorithm showing the edges in the
$\lambda$CDM map of Figure 2.} 
\label{fig:4}
\end{figure}

\begin{figure}
\includegraphics[height=9cm]{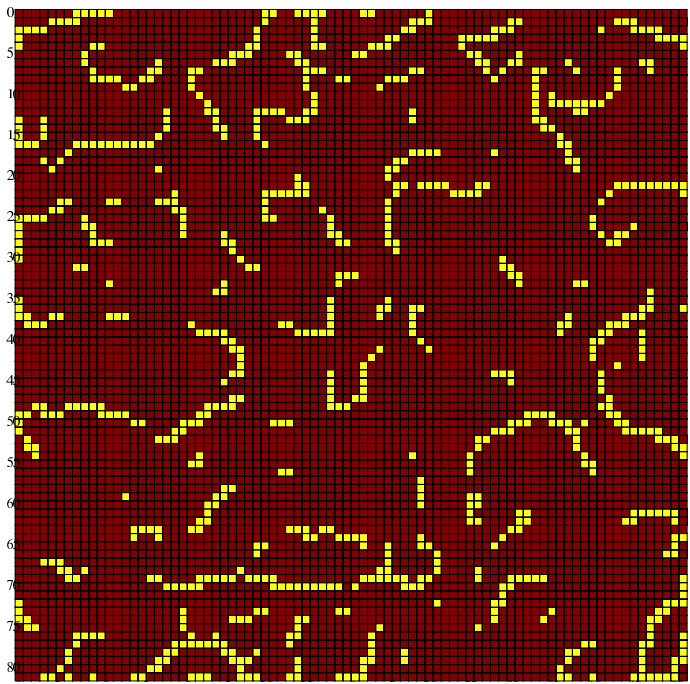}
\caption{Corresponding edge distribution in the map of Figure 3.} \label{fig:5}
\end{figure}

Based on these results, our algorithm evaluates the distribution of
edge lengths and checks whether the distributions of the histograms
with and without strings are statistically different. 
Maps with cosmic strings showed, as expected,
a slight excess of edges of all lengths. This excess is more
pronounced for edge lengths which lie in the non-Gaussian tail. 

The results for an
angular resolution of $8^{'}$ are presented in Table 1. In this
table, the rows indicate the length $L$ of the edge. The entries
in the columns are the mean number [standard deviation of the mean] 
of edges of the maps which have the respective length. The
first column of numbers is for simulations without strings, the
next two are for simulations including a network of strings with the
indicated mass per unit length parameter. Each histogram
was produced from 50 runs. 
From the histogram it follows that the difference between
the distributions is significant (at the $3 \sigma$ level) 
if $G \mu = 3.5 \times 10^{-8}$,
but not if $G \mu = 3 \times 10^{-8}$.

For a given angular resolution, we varied the string mass parameter 
$G \mu$ to find the limiting value below
which the differences in the histograms ceased to be statistically
significant (using Fisher's combined probability test). 
Our results are summarized in Table 2.

Based on the above results, prospects for being able to use the
Canny algorithm to significantly improve the limits on the
cosmic string mass parameter $G \mu$ (or detect cosmic strings)
are excellent. The angular resolution of the Planck satellite
experiment will be about $5^{'}$, the South Pole Telescope is
aiming for a resolution of $1^{'}$ with a field of view of
$4,000$ square degrees, and the resolution of the ACT telescope
in Chile will be $1.5^{'}$. 

\begin{table}
\caption{Histograms of the distribution of edge lengths $L$.}
\label{table:1}
\begin{center}
\begin{tabular}{cccc} \hline \hline
& no strings & $G \mu = 3 \times 10^{-8}$ & $G \mu = 3.5 \times 10^{-8}$   
\\ \hline
L = 2 &  165.0 [2.84]   &  169.9 [3.60] &  178.0 [3.84] \\
L = 3 &  40.0 [0.86]     &  42.3 [1.19]   &  43.6 [1.18] \\
L = 4 &  11.8 [0.51]     &  13.4 [0.58]   &  14.0 [0.57] \\
L = 5 &  4.5 [0.31]     &   4.8 [0.32]   &  5.5 [0.34] \\
L = 6 &  1.7 [0.23]     &   2.1 [0.21]   &  2.2 [0.21] \\
L = 7 &  0.66 [0.13]     &  0.74 [0.11]   &  0.84 [0.12] \\
L = 8 &  0.18 [0.07]     &  0.38 [0.09]   &  0.28 [0.07] \\
L = 9 &  0.12 [0.06]     &  0.1 [0.04]   &  0.08 [0.05] \\
L = 10 &  0.04 [0.03]     &  0.08 [0.04]   &  0.04 [0.03] \\ \hline
\end{tabular}
\end{center}
\end{table}

\begin{table}
\caption{Critical values of the string mass parameter $G \mu$.}
\label{table:2}
\begin{center}
\begin{tabular}{ccc} \hline \hline
angular resolution & $G \mu$   \\ \hline
$10^{'}$     & $4.3 \times 10^{-8}$ \\
$9^{'}$      & $4.0 \times 10^{-8}$ \\
$8^{'}$      & $3.1 \times 10^{-8}$ \\ \hline
\end{tabular}
\end{center}
\end{table}

\section{Conclusions}
 
We have suggested a new way of looking for the specific signature
of cosmic strings, namely the Kaiser-Stebbins
line discontinuities, in small-scale 
cosmic microwave anisotropy maps. Our
method makes use of the Canny algorithm, an edge-detection technique
used often in pattern recognition.

Since the cosmic microwave temperature anisotropy maps induced by
a network of cosmic strings are dominated by the strings present at
the time of last scattering (when the Hubble radius, which is
comparable to the correlation length of the string network at that
time, is of the order of one degree), good small-scale angular resolution 
is essential in order to be able to detect strings. An angular resolution
substantially less than $1^{o}$ is required, otherwise the signals
from the line discontinuities are washed out. 

We have constructed CMB anisotropy maps which correspond to having
both Gaussian "noise" with a nearly scale-invariant power spectrum
from inflation and anisotropies produced by a distribution of
straight string segments. Based on our numerical simulations, 
we find that for an angular resolution
of CMB maps of $8^{'}$ the Canny algorithm has the potential
to detect strings with a mass per unit length $\mu$ above a value
of $G \mu \simeq 4 \times 10^{-8}$, close to an order of magnitude
better than current limits based on the CMB.

A drawback of our work is that it is based on toy model cosmic string
CMB maps which are obtained by superimposing idealized line
discontinuities of straight string segments, not from actual
string networks. Actual string networks contain both infinite strings
and string loops. The infinite strings are not completely straight,
but have a curvature radius comparable to the Hubble radius. 
An improved analysis should start from a numerical simulation of the 
distribution of cosmic strings, calculate the induced temperature 
anisotropies taking into account of all effects following the formalism 
set out in \cite{Turok},
and then apply the Canny algorithm to the resulting maps. After
completion of this work, a paper appeared \cite{Fraisse} in which
CMB temperature maps on scales similar to the ones we are considering
were constructed based on full cosmic string simulations. This
work confirmed that the Kaiser-Stebbins from long string segments
is the dominating visible effect, and suggested, in agreement with
the message of our work, that strings should be visible in high
resolution small-scale CMB anisotropy maps. However, no string-specific
statistical analyses of the maps like the one we are proposing were
performed in \cite{Fraisse}. 

Another important issue which remains to be analyzed is the effect
of instrumental noise. We have done preliminary work on this
topic and modelled instrumental noise by a component to the $C_l$
spectrum which rises rapidly as a function of $l$ at the scale of
the angular resolution of the survey \footnote{We thank Gil Holder
for suggesting this method.}, becoming dominant at the
angular resolution scale. Initial results show that the efficiency
of the Canny algorithm is not reduced. We plan to study this question
in more detail.

\begin{acknowledgments} 
 
This work is supported in part by a NSERC Discovery Grant, by funds from
the CRC Program, and by a FQRNT Team Grant. We wish to thank Matt Dobbs,
Christophe Ringeval, and in particular Gil Holder for useful discussions.

\end{acknowledgments}

\end{document}